\documentstyle[12pt]{article}

\begin{document}


\baselineskip=4mm

\title{\bf The h-index and the Harmonic Mean}
\author{M. Germano  \\ \\
\small Department of Civil and Environmental Engineering\\
\small Duke University, Durham, NC 27708, USA\\
\small mg234@duke.edu}

\date{}
\maketitle


\abstract{The Harmonic Mean between the number of papers and the citation number per paper is proposed as a simple single-value index to quantify an individual's research output. Two simple comparisons with the Hirsch h-index are performed.}

\section*{}
The single-number criteria commonly used to evaluate scientific output of a researcher are the number of papers $N_{p}$, a rough measure of the {\it productivity}, and the total number of citations $N_{c,tot}$, a rough measure of the {\it impact}. A third simple single-number index is the ratio between $N_{c,tot}$ and $N_{p}$, the citations per paper $N_{c}$. Advantages and disadvantages of these simple single-numbers indices are discussed in \cite{Hir05}, where the celebrated h-index is proposed: {\it A scientist has index h if h of his/her $N_{p}$ papers have at least h citations each, and the other ($N_{p}$ - h) papers have no more than h citations each}. One basic motivation was to avoid unjustified reward or penalization of low productivity and low impact. The merits of the Hirsch h-index are provided by an academic example. Let us consider the lives of five researchers characterized scientifically by the following parameters
\begin{center}
\begin{tabular}{|c|c|c|c|}
  \hline
   Name & $N_{p}$ & $N_{c,tot}$ & $N_{c}$ \\
  \hline
   Researcher 1 & 1 & 10000 & 10000  \\
   Researcher 2 & 10 & 10000 & 1000  \\
   Researcher 3 & 100 & 10000 & 100  \\
   Researcher 4 & 1000 & 10000 & 10  \\
   Researcher 5 & 10000 & 10000 & 1  \\
  \hline
\end{tabular}
\end{center}
We remark that they have the same number of citations but very different productivity and impact, and their h-indices are given by
\begin{center}
\begin{tabular}{|c|c|c|c|c|}
  \hline
   Name & $N_{p}$ & $N_{c,tot}$ & $N_{c}$ & h\\
   \hline
   Researcher 1 & 1 & 10000 & 10000 & 1  \\
   Researcher 2 & 10 & 10000 & 1000 & 10  \\
   Researcher 3 & 100 & 10000 & 100 & 100  \\
   Researcher 4 & 1000 & 10000 & 10 & 10  \\
   Researcher 5 & 10000 & 10000 & 1 & 1  \\
  \hline
\end{tabular}
\end{center}

Strangely enough the merits of another simple index have not been explored in detail, at least to the knowledge of the author. Let us consider the half Harmonic Mean between the productivity, quantified by the number of papers $N_{p}$, and the impact, quantified by the citations per paper $N_{c}$
\begin{equation}\label{eq:id1}
\frac{1}{H}=\frac{1}{N_{p}}+\frac{1}{N_{c}}
\end{equation}
If we compare in our academic example the HM-index with the Hirsch h-index  we have
\begin{center}
\begin{tabular}{|c|c|c|c|c|c|}
  \hline
   Name & $N_{p}$ & $N_{c,tot}$ & $N_{c}$ & h & HM\\
   \hline
   Researcher 1 & 1 & 10000 & 10000 & 1 & 1 \\
   Researcher 2 & 10 & 10000 & 1000 & 10  &  10\\
   Researcher 3 & 100 & 10000 & 100 & 100 & 50 \\
   Researcher 4 & 1000 & 10000 & 10 & 10 & 10 \\
   Researcher 5 & 10000 & 10000 & 1 & 1 & 1 \\
  \hline
\end{tabular}
\end{center}
and we see that the $HM$-index agrees with the $h$-index to penalize outrageous productivity and impact, but is a little less disposed to reward the middle.

Another example is more realistic. Thirty years ago three researchers gathered at the Center of Turbulent Research in Stanford and elaborated with the Professor Parviz Moin the so called dynamic model, a subgrid procedure useful in the Large Eddy Simulation of turbulent flows \cite{Ger91}. They were Massimo Germano, Ugo Piomelli and William H. Cabot. Since then they dispersed in the world, the first is now at Duke University, the second at Queen's University and the last at the Lawrence Livermore National Laboratory. The Professor Moin is always at the CTR, and during this long time everyone developed his own life of researcher. They are very interesting from the point of view of the quantification of the research. From Scopus, we derive on September $30$, $2020$ the following data
\begin{center}
\begin{tabular}{|c|c|c|c|}
  \hline
   Name & $N_{p}$ & $N_{c,tot}$ & h \\
   \hline
   Germano & 37 & 6235 & 9 \\
   Piomelli & 150 & 11467 & 39 \\
   Moin & 288 & 38042 & 86 \\
   Cabot & 39 & 9128 & 21 \\
  \hline
\end{tabular}
\end{center}
We see that two of them have a large number of papers and citations, the other two a relatively large number of citations but a lower productivity. The $h$-index penalizes them severely. The Harmonic Mean $HM$-index is more generous. We have
\begin{center}
\begin{tabular}{|c|c|c|c|c|c|}
  \hline
   Name & $N_{p}$ & $N_{c,tot}$ & $N_{c}$ & h & HM\\
   \hline
   Germano & 37 & 6235 & 168 & 9 & 30 \\
   Piomelli & 150 & 11467 & 76 & 39 & 51 \\
   Moin & 288 & 38042 & 132 & 86 & 91 \\
   Cabot & 39 & 9128 & 234 & 21 & 33 \\
  \hline
\end{tabular}
\end{center}

\end{document}